\documentclass[%
 aps,
 amsmath,amssymb,
 reprint,%
]{revtex4-2}

\usepackage{graphicx}
\usepackage{subfig}
\usepackage{dcolumn}
\usepackage{bm}
\usepackage{amsmath}
\usepackage[dvipsnames]{xcolor}
\usepackage{orcidlink}%

\begin{document}

\preprint{PRD}

\title[Planck-Scale Spacetime Lattice]{A Lattice Physics Approach to Spin-Networks in Loop Quantum Gravity}

\author{Noah M. MacKay \,\orcidlink{0000-0001-6625-2321}}
 \email{noah.mackay@uni-potsdam.de}
\affiliation{%
Institut für Physik und Astronomie, Universität Potsdam\\
Karl-Liebknecht-Straße 24/25, 14476 Potsdam, Germany
}%

\date{\today}

\begin{abstract}
In this study, we model a spin-network in loop quantum gravity as a regular tetrahedral lattice, applying lattice physics techniques to study its structure and vertex dynamics. Using the area eigenvalue, $A\propto 8\pi l_P^2$, we derive a lattice constant $a = 2.707\,l_P$ and construct a vertex Hamiltonian incorporating a Lennard-Jones potential, zero-point energy, and simple harmonic oscillations. A foliation approach enforces the Wheeler-DeWitt constraint via locally non-zero Hamiltonians that globally cancel. Graviton-like perturbations (treated here as spin-0 bosons) modify the vertex energy spectrum, with variational analysis suggesting twelve coherent excitations per vertex. This model frames flat spacetime as a graviton-rich lattice while enforcing a Brownian-like stochastic picture for the gravitons, and offers a basis for extension into curved quantum geometries.
\end{abstract}

\maketitle

\section{Introduction}

One of the central challenges in modern physics is the unification of general relativity and quantum mechanics into a theory of quantum gravity. Since Feynman's pioneering work using effective field theory \cite{Feynman:2002, Rafie-Zinedine:2018izq}, this pursuit has inspired multiple theoretical frameworks, including string theory \cite{Duff:1996aw, Becker:2007}, stochastic gravity \cite{Hu:1994ep, Moffat:1996fu}, and loop quantum gravity (LQG) \cite{Rovelli:1987df, Rovelli:1989za, Rovelli:1994ge}.   

LQG offers a fundamentally geometric approach, treating spacetime itself as a quantized structure, akin to Einstein's interpretation of gravity as the response effect to curved spacetime geometry \cite{Einstein:1915ca}. This stands in contrast to approaches that attempt to quantize gravity as a force mediated by a massless spin-2 particle: the graviton \cite{Feynman:2002, Rafie-Zinedine:2018izq, Duff:1996aw, Becker:2007, MacKay:2024qxj, MacKay:2024owk, MacKay:2024sgw}. 

In LQG, spacetime is composed of discrete quanta woven into a spin-network: a graph-like structure encoding quantum states of geometry \cite{Rovelli:1994ge}. This spin-network admits an area operator; when interpreted as a fastened yet vibrating lattice \cite{Rovelli:2007}, its area eigenvalues take the form
\begin{equation} \label{lqga}
A=8\pi G\hbar\gamma_I\sum_i \sqrt{j_i(j_i+1)}.
\end{equation}
Here, $G\hbar$ defines the square of the Planck length ($l_P\equiv\sqrt{G\hbar}=1.6\times10^{-35}~\text{m}$,  with $c=1$), $\gamma_I$ is the dimensionless Immirzi parameter, and $j_i$ are half- or whole-integer spin numbers labeling the spin-network edges \cite{Rovelli:1994ge, Hedrich:2009mm}. Under a lattice approximation, the lattice spacing between adjacent quanta (the lattice constant $a$) can be related to the area eigenvalue through a geometric scaling relation, with surface area $A_S\propto a^2$.

When accounting for time evolution, spatial spin-networks evolve into spin-foams, which describe the quantum dynamics of local spacetime regions. The Wheeler-DeWitt equation \cite{DeWitt:1967yk},
\begin{equation}
\widehat{\mathcal{H}}|\Psi\rangle = 0,
\end{equation} 
imposes the Hamiltonian constraint, requiring that physical spin-network states $|\Psi\rangle$ be annihilated by the quantum Hamiltonian. This constraint reflects the underlying diffeomorphism invariance and the absence of an external time parameter in general relativity. LQG, formulated on the Arnowitt-Deser-Misner (ADM) formalism \cite{Arnowitt:1962hi}, foliates spacetime into spatial hypersurfaces, each described by a spin-network. The Hamiltonian constraint ensures consistent quantum evolution between these slices, with the total constraint vanishing for all physical states. 

Building on the area spectrum and the Wheeler-DeWitt constraint, we aim to characterize a single spin-network modeled as a Planck-scale spacetime lattice under simplified assumptions. While LQG describes spatial geometry via spin-networks and time evolution via spin foams \cite{Rovelli:1994ge}, the treatment of time within a static spin-network remains an open issue. Accordingly, we restrict our attention to three-dimensional representation. We propose to investigate the following: 

\begin{itemize}
\item The effective interaction between two neighboring quanta in the lattice, modeled using a Lennard-Jones potential, with each quantum assumed to carry a spherical harmonic azimuthal number $l$.
\item The quantization of vibrational modes in a spin-network, interpreted as graviton-like excitations in the semi-classical limit (i.e., gravitons as emergent pseudoparticles).
\item The behavior of these vibrational modes as a free bosonic gas within the lattice, analogous to spin-0 phonons in condensed matter systems (c.f. Ref. \cite{Hu:2005ub}).
\end{itemize}

\subsection{3D Representation of a Spacetime Lattice} \label{3drep}

To approximate a three-dimensional spin-network in LQG, we adopt a tetrahedral lattice representation that preserves the triangulation structure inherent in the two-dimensional model \cite{Rovelli:1994ge, Rovelli:2007}. In the 2D case, nodes represent quanta of volume and edges represent quanta of area, with each node connected to six neighbors forming triangular plaquettes. Extending this structure to three dimensions, a tetrahedral lattice comprises vertices connected by edges that form tetrahedral cells. Each vertex is coordinated with six neighbors, maintaining a triangulated and homogeneous structure.

\begin{figure}[h!]
\centering
\includegraphics[width=0.33\textwidth]{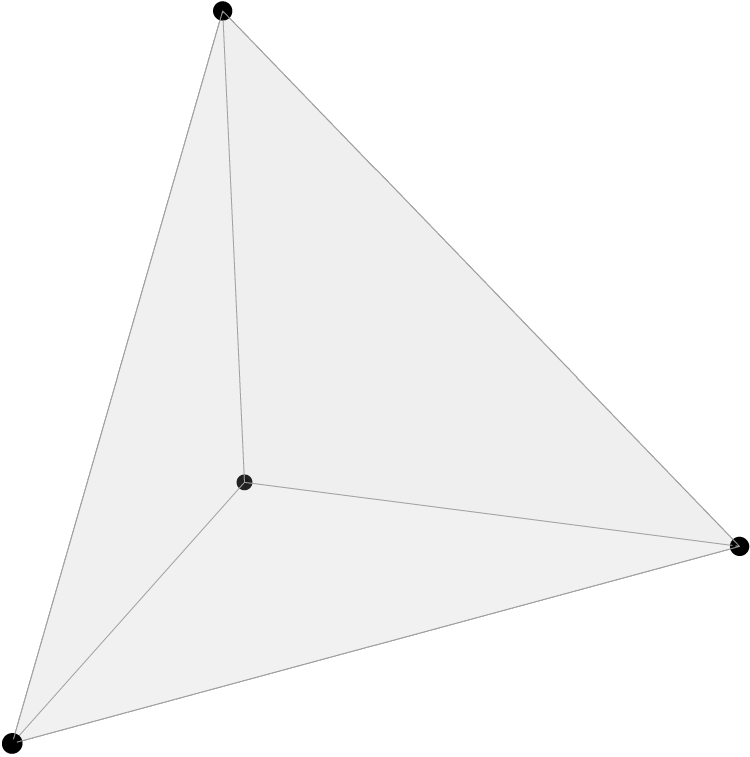}
\caption{A triangular pyramid (tetrahedron) as a Delaunay mesh lattice cell.}\label{fig:sub}
\end{figure}

Each tetrahedral cell consists of four vertices (lattice points), four triangular faces, and six edges, as shown in Figure \ref{fig:sub}. If any one of the four faces is taken as the base, with the base length identified as the lattice constant $a$, the surface area of a tetrahedron is given by the following (see Ref. \cite{Weisstein:2005}):
\begin{equation}
 A_\text{tetra}=\frac{\sqrt{3}}{4}a\left(a+\sqrt{a^2+12h^2}\right),
\end{equation}
where $h$ is the height from the apex to the base face. By geometric construction, the half-distance of a median is $a\sqrt{3}/4$; applying the Pythagorean theorem then yields the tetrahedral height $h=a\sqrt{13}/4$ (see Appendix \ref{calc} for derivation). Substituting this result into the surface area expression produces a closed-form formula in terms of the lattice constant:
\begin{equation}\label{tetra}
 A_\text{tetra}=\frac{\sqrt{3}}{4}\left(1+\frac{\sqrt{43}}{2}\right)a^2.
\end{equation}

We adopt the following assumptions:

\begin{itemize}
\item Each vertex has six neighbors: three forming the reference tetrahedron structure, and three belonging to adjacent cells in which the vertex is also included.
\item Lattice edges do not intersect or ``cross-thread.''
\end{itemize}

The latter assumption preserves the combinatorial graph structure of a spin-network, where nodes serve as junctions in a non-intersecting graph (c.f. Ref. \cite{Thiemann:2007pyv}). This regular tetrahedral lattice serves as an idealized simplification of the generally irregular spin-network topology in LQG. It enables tractable analysis of local interactions, such as those modeled by effective potentials, and dynamical phenomena such as vibrational modes.

Using the area eigenvalue formula from Eq. (\ref{lqga}), we assign the upper summation limit to 6 edges, corresponding to the six edges of a tetrahedral cell:
\begin{equation}\label{equiv1}
8\pi l_P^2\gamma_I\sum_{i=1}^{6} \sqrt{j_i(j_i+1)}=\frac{\sqrt{3}}{4}\left(1+\frac{\sqrt{43}}{2}\right)a^2.
\end{equation}
We make the simplifying assumption that the spin numbers $j_i$ are fermionic, motivated by the interpretation that fermionic assignments correspond to flat spacetime configurations \cite{Hedrich:2009mm}. For a minimal lattice model, we set all spins to the lowest, non-zero value of $j_i=1/2$. Additionally, the Immirzi parameter for a spin-network is taken as $\gamma_I=\ln(3)/(\pi\sqrt{8})\approx3/25$ \cite{Ansari:2006vg, Ansari:2006cx}. Substituting these values, the surface area evaluates to
\begin{equation}
8\pi l_P^2\frac{18}{25} \sqrt{\frac{1}{2}\left(\frac{1}{2}+1\right)}= 13.572~l_P^2,
\end{equation}
which, via Eq. (\ref{equiv1}), yields a lattice constant $a=2.707\,l_P$.

\section{The Hamiltonian of One Lattice Point}

The Hamiltonian constraint, encoded in the Wheeler-DeWitt equation, ensures that a local region of spacetime is diffeomorphism-invariant and locally flat in the quantum geometric sense. We propose that this global constraint emerges from a sum over local Hamiltonian operators defined on our tetrahedral lattice. Each local Hamiltonian corresponds to a spatial slice (in the sense of a foliation), and contributes a non-zero curvature term in the form of quantum dynamics. These contributions collectively cancel, satisfying the global constraint.

\begin{figure}[h!]
\centering
\includegraphics[width=0.4\textwidth]{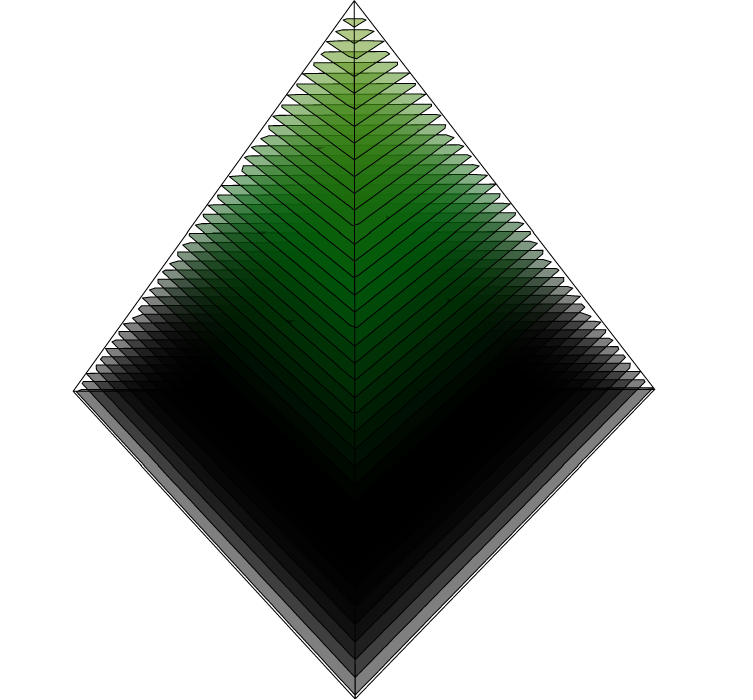}
\caption{A pictorial representation of a foliated tetrahedral cell. The bottom (black) slice is tangent to the bottom face, and the top (green) slice converges to the apex.}\label{fig:slice}
\end{figure}

Focusing on a single tetrahedral cell, we define a foliation by partitioning the cell into $M$ spatial slices: the bottom slice contains the base face $f_\text{b}$, the top slice contains the apex vertex $v_\text{a}$, and the intermediate slices (indexed by $m=2,\dots,M-1$) consist of subgraphs with edges in the interior (bulk). The global Hamiltonian constraint can be expressed as
\begin{equation}
\widehat{\mathcal{H}}|\Psi\rangle=\sum_{n=1}^M \widehat{H}_n|\psi\rangle_n=0,
\end{equation}
where $|\Psi\rangle$ denotes the global spin-network state, $|\psi\rangle_n$ the state on the $n$-th slice, and $\widehat{H}_n$ the corresponding local Hamiltonian operator. Each $\widehat{H}_n$ is constructed from holonomies and fluxes associated with the slice's edges and vertices, following standard LQG quantization procedures \cite{Thiemann:1996aw}. A visual aid for this decomposition is shown in Figure \ref{fig:slice}.

We define the Hamiltonians on the apex (top) and base (bottom) slices as
\begin{equation}
\widehat{H}_\text{t}|\psi\rangle_\text{t}=\widehat{H}(v_\text{a})|\psi\rangle_\text{a},\quad\quad\widehat{H}_\text{b}|\psi\rangle_\text{b}=\widehat{H}(f_\text{b})|\psi\rangle_\text{b},
\end{equation}
where $\widehat{H}(v)$ and $\widehat{H}(f)$ denote vertex and face contribution to the Hamiltonian constraint, incorporating the volume operator and curvature of the Ashtekar-Barbero connection. For each intermediate slice, indexed as $m=2,\dots,M-1$, the local Hamiltonain is defined as a sum over the contributions of all edges and the bulk intersecting that slice:
\begin{equation}
\widehat{H}_m=\sum_{e\in\text{slice}~m}\widehat{H}(e).
\end{equation}
The full Wheeler-DeWitt constraint across the foliated lattice cell then reads:
\begin{equation}
\widehat{H}_\text{t}|\psi\rangle_\text{t}+ \widehat{H}_\text{b}|\psi\rangle_\text{b}+ \sum_{m=2}^{M-1}\widehat{H}_m|\psi\rangle_m=0.
\end{equation}

Due to a combinatorial symmetry in the tetrahedron (specifically, a duality between the apex vertex $v_\text{t}$ and the triplet of vertices and edges that form the base face $f_\text{b}$), the corresponding top and bottom Hamiltonians satisfy a reflection relation:
\begin{equation}\label{reflect}
\widehat{H}_\text{t}|\psi\rangle_\text{t}=- \widehat{H}_\text{b}|\psi\rangle_\text{b}.
\end{equation}
This antisymmetry, combined with the vanishing contribution of intermediate slices:
\begin{equation}
\sum_{m=2}^{M-1}\widehat{H}_m|\psi\rangle_m=0,
\end{equation}
ensures that the localized hyperspatial curvatures cancel out, satisfying the flatness condition of the global Hamiltonian constraint. This foliation scheme mimics the ADM decomposition of spacetime into spatial hypersurfaces \cite{Arnowitt:1962hi}, with each slice encoding its own quantum geometry and contributing to local dynamics, while the full structure maintains diffeomorphism invariant and vanishing curvature in the spin-network background.

Given the reflection relation in Eq. (\ref{reflect}), we may equivalently consider either the top slice containing a single resident vertex $v_\text{t}$, or the bottom slice containing the face $f_\text{b}$, which comprises three vertices, three edges, and a vacuum region supporting vibrational modes. For simplicity, we focus on the top slice.

We model the associating Hamiltonian with a kinetic term augmented by an angular momentum barrier to account for geometric structure, a cohesive potential energy incorporating an effective Lennard-Jones interaction with one of its six neighbors, a quantum zero-point energy, and a vibrational potential term:
\begin{equation}\label{ham}
\widehat{H}(v_1)=-\frac{\hbar^2}{2m^*}\left(\nabla^2-\frac{l(l+1)}{r^2} \right)+V_\text{cohes.}+V_\text{vibr.}.
\end{equation} 
Here, $m^*$ is the effective mass of the lattice point. We estimate $m^*$ via its associating Compton wavelength $\lambda_C=h/m^*$, assuming the lattice point's influence extends across a characteristic radius $r_c=a/2=1.354\, l_P$. If we take the equatorial standing wavelength of this spherical cell to represent $\lambda_C$, then $\lambda_C=2\pi r_c$, yielding $m^*=0.7388\,m_P$, where $m_P\equiv\sqrt{\hbar/G}=2.17\times10^{-8}~\text{kg}$ is the Planck mass.

\subsection{Lattice Structure Energy}

Given Eq. (\ref{ham}), the Laplacian (free kinetic) term together with the vibrational potential energy defines a simple harmonic oscillator (SHO) contribution $U_\text{SHO}$ that describes the vibrational energy from the lattice points. Meanwhile, the angular momentum term combines with the cohesive energy to form a structural energy $U_s$ associated with the interaction between two adjacent lattice points. We model this structural energy using the angular momentum barrier with azimuthal number $l$, a Lennard-Jones (12-6) potential \cite{Lennard-Jones:1931, Lenhard:2024}, and a zero-point quantum energy:
\begin{equation}\label{structure}
U_s(r)=\frac{\hbar^2}{2m^*}\frac{l(l+1)}{r^2}+U_\text{ZP}+4\epsilon\left[\left(\frac{\sigma}{r}\right)^{12}-\left(\frac{\sigma}{r}\right)^6\right].
\end{equation}
Here, $r$ is the radial separation between the lattice points, $\sigma$ is the Lennard-Jones null point: i.e., the point of zero potential, $\epsilon$ is the depth of the potential well, and $U_\text{ZP}=(2\pi\hbar/\sigma)\sqrt{\epsilon/m^*}$ is the zero-point quantum energy associated with oscillations at scale $\sigma$. 

To determine the null point $\sigma$, we minimize the structural energy in Eq. (\ref{structure}) by setting the derivative with respect to $r$ equal to zero at equilibrium $r_0$:
\begin{equation} \label{simp}
\partial_r U_s(r)\Big|_{r=r_0}=0~\implies~x^4-2x^{10} =\frac{\alpha}{12\epsilon},
\end{equation}
where $x=\sigma/r_0$ and $\alpha=\hbar^2l(l+1)/(2m^*\sigma^2)$. This yields a polynomial equation in $x\propto \sigma$, whose roots determine the null point $\sigma$ for a given equilibrium point $r_0$.  However, the solutions to Eq. (\ref{simp}) depend on the ratio $\alpha/\epsilon$; we simplify the analysis by choosing $\alpha=\epsilon$, which yields four real roots: $x=\pm0.8676,~\pm0.5445$. Taking the positive values, we find $\sigma=0.8676r_0$ and $\sigma=0.5445r_0$, corresponding to two possible lengths. 

For comparison, the minimization of the standard 12-6 potential (i.e., a potential with no angular momentum term) gives $x=2^{-1/6}\approx0.8909$, or equivalently $\sigma=0.8909r_0$. This provides a useful benchmark: our solution $\sigma=0.8676r_0$ lies close to the classical result, indicating that the angular momentum barrier introduces only a small correction to the equilibrium structure. 

Physically, the equilibrium point $r_0$ marks the balance between the short-range Pauli repulsion (scaling as $\sim1/r^{12}$) and  the longer range dipole attraction (scaling as $\sim-1/r^6$) \cite{Lenhard:2024}. Assuming this balance occurs at the midpoint of a lattice edge, $r_0=a/2=1.3535\, l_P$, we obtain $\sigma=1.1743\,l_P$. We can now solve for the Lennard-Jones well depth $\epsilon$ under the working ansatz $\epsilon=\alpha$. Recovering $c$, we obtain:
\begin{equation}
\epsilon=0.4908\,{m_P c^2}\,l(l+1),
\end{equation}
and the corresponding zero-point quantum energy becomes:
\begin{equation}
U_\text{ZP}=4.3610\,m_Pc^2\sqrt{l(l+1)}.
\end{equation}
It is crucial to note that both $\epsilon$ and $U_\text{ZP}$ vanish when azimuthal number $l$ is zero. That is, if $l=0$, the entire structural energy collapses to zero. This strongly suggests that a quantum spacetime lattice fundamentally requires non-zero angular structure: not just in the edge spin number $j_i$, but also at the vertex level through angular momentum excitations. These angular degrees of freedom (whether from $j_i$ or $l$) point toward a deeper necessity of a spin-foam framework, in which vertex-centered angular excitations contribute to the fabric of spacetime itself. The angular momentum-dependent structural energy behaves like an emergent Brownian potential landscape in a canonically flat spacetime (further supporting \cite{Hu:1994ep, Moffat:1996fu}), embedding quantum fluctuations through geometry.

\begin{figure}[h!]
\centering
\includegraphics[width=0.46\textwidth]{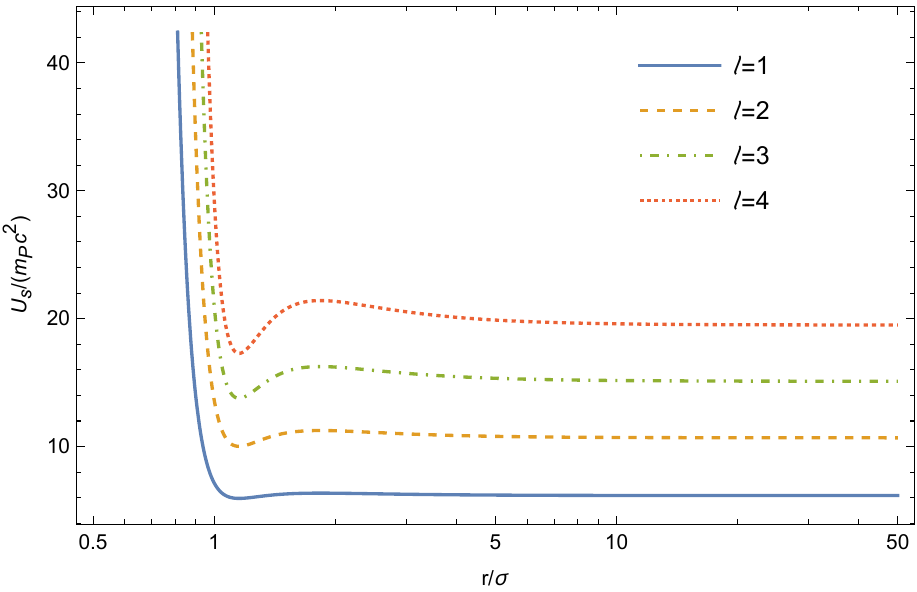}
\caption{Normalized structural energy $U^{(l)}_s(r)/(m_Pc^2)$ plotted as a function of $r/\sigma$, for azimuthal numbers $l=1$ through $4$. Higher $l$-modes increase the overall energy scale, deepen the Lennard-Jones (12-6) potential well, and steepen the centrifugal barrier.}\label{fig:stru}
\end{figure}

The full angular momentum dependent structural energy, expressed as a polynomial of $\sigma/r$, becomes:
\begin{widetext}
 \begin{equation}\label{fstru}
 U^{(l)}_s(r)=m_Pc^2~l(l+1)\left\{\frac{4.3610}{\sqrt{l(l+1)}}+0.4908\left(\frac{\sigma}{r}\right)^2+1.9632\left[\left(\frac{\sigma}{r}\right)^{12}-\left(\frac{\sigma}{r}\right)^6\right]\right\}.
 \end{equation}
\end{widetext}
For the lowest non-zero azimuthal number, $l=1$, the structural energy $U_s^{(1)}(r)$ takes on characteristic values of $7.149\,m_Pc^2$ at the null point $r=\sigma$, $5.9459\,m_Pc^2$ at the equilibrium point $r=r_0$, and $6.1674\,m_Pc^2$ as $r\rightarrow\infty$. Figure \ref{fig:stru} presents the normalized structural enrgy $U^{(l)}_s(r)/(m_Pc^2)$ as a function of the radial ratio $r/\sigma$, plotted for azimuthal numbers $1\leq l\leq4$.

\subsection{Spin-Network Fluctuations via $U_\text{SHO}$}

We now consider the simple harmonic oscillator (SHO) energy $U_\text{SHO}$, constructed from the Laplacian (free kinetic) term and the vibrational potential energy in the vertex Hamiltonian. This energy admits the familiar quantization of harmonic oscillations: 
\begin{equation}
U_\text{SHO}=\sum_{\vec{k},s}\hbar\omega_s(\vec{k})\left(n_s+\frac{1}{2}\right),
\end{equation}
where $\omega_s(\vec{k})$ denotes the angular frequency of the vertex oscillations as a function of the wave number vector $\vec{k}$, and $n_s$ indexes the quantized energy states.

In analogy with condensed matter systems, where phonons are the quanta of vibrational modes, these oscillations on a spacetime lattice correspond to effective gravitational excitations. Motivated by stochastic gravity frameworks \cite{Hu:1994ep, Moffat:1996fu}, we interpret such fluctuations as gravitons emerging from a Brownian-bath-like background of metric perturbations. However, within LQG, gravitons are not fundamental particles but pseudoparticles, emergent from coarse-grained quantum geometric states. This perspective is particularly applicable in our model, which restricts to a three-dimensional spatial representation without an explicit time direction. The usual spin-2 characterization of gravitons arises from their tensorial role in 3+1 dimensional spacetime; alternative intepretations attribute the graviton's spin to its coupling via the stress-energy tensor in the Einstein field equations \cite{Mogull:2020sak}. In a purely spatial model, that interpretation does not strictly apply. Therefore, in this setting, we treat gravitons as spin-0 pseudoparticles with a degeneracy factor $g_s=1$. 

Since gravitons, whether fundamental or pseudoparticle-like, are bosons, their state occupation is governed by Bose-Einstein statistics. If these bosons are coherent, they predominantly occupy a single energy state, replacing the SHO energy state $n_s$ with the Bose-Einstein distribution $\bar{n}_\text{BE}$. The SHO energy then becomes:
\begin{equation}
U_\text{SHO}=\hbar\sum_{\vec{k},s}\omega_s(\vec{k})\left(\frac{1}{\exp(\hbar\omega_s(\vec{k})/k_BT)-1}+\frac{1}{2}\right),
\end{equation}
where $k_BT$ is the thermal energy scale.

As massless quanta of SHO vibrations, gravitons propagate at the speed of light. In natural units ($c=1$), their dispersion relation becomes linear: $\omega=|\vec{k}|$. Since the SHO energy is defined in $k$-space, we perform an inverse Fourier transformation to express the energy in physical space, where the inverse-length dimensionality is normalized using the lattice constant $a$. The resulting SHO energy in position space is
\begin{equation}\label{sho}
U_\text{SHO}(r)=-\frac{(k_BT)^2}{\hbar}a\sqrt{\frac{\pi^3}{2}}\mathrm{csch}^2\left(\frac{k_BT}{\hbar}\pi r\right).
\end{equation}
The physical (real-valued) energy corresponds to the real component of the transformation. The imaginary part, $i\hbar\partial_r\delta(r)\sqrt{\pi/2}$, contains a spatial derivative on the Dirac delta function and is discarded on physical grounds. The negative sign in Eq. (\ref{sho}) arises from the inversion procedure but also admits a physical interpretation: the SHO energy in real space reflects diffusive behavior, characteristic of coherent bosonic systems in fluctuating (Brownian-like) backgrounds (see e.g. Ref. \cite{MacKay:2024owk}).

Under thermal equilibrium, the system and its environment share a common temperature $T$. In the vacuum of space, this background temperature is approximately $T=2.725$ K, set by the cosmic microwave background (CMB). Such low temperatures would, in a free gas model, suppress the Brownian motion of graviton-like excitations. However, in the presence of matter (here, the spacetime quanta consisting the lattice vertices), gravitons remain dynamically active due to their mutual attraction to energy-momentum sources. That is, while thermally suppressed in empty space, they still fluctuate stochastically in the presence of geometry. To describe this behavior, we must introduce an effective temperature $T^*$, characterizing the thermal energy scale at which gravitons exhibit stochastic dynamics within the spacetime lattice. This effective temperature is not defined by the CMB, but by internal lattice properties.

We estimate $T^*$ by minimizing the physical-space SHO energy towards the equilibrium point $r_0=a/2$; this gives the following extremum condition:
\begin{widetext}
\begin{equation}
\partial_{r}U_\text{SHO}(r)\Big|_{r=r_0}=\sqrt{2\pi^5}\frac{(k_BT^*)^3}{\hbar^2}a\coth\left(\frac{k_BT^*}{2\hbar}\pi a\right)\mathrm{csch}^2\left(\frac{k_BT^*}{2\hbar}\pi a\right)=0.
\end{equation}
\end{widetext}
Defining $k_BT^*(\pi a)/(2\hbar)=:y$, the minimization condition effectively reads as $y^3\coth(y)\mathrm{csch}^2(y)=0$. This function does not provide exact zeroes, as it has a profile of a flat-head bell curve that decays rapidly but asymptotically towards zero only in the limit $y\rightarrow\infty$. Nonetheless, by setting a practical threshold for numerical smallness (e.g., using \texttt{FindRoot} in \textsc{Wolfram Mathematica} for $y\in[1,\,500]$), we can approximate $y$ based on a desired truncation order. E.g., for a modest truncation at $10^{-5}$, we yield $y=9.88634\approx \pi^2$; at a tighter truncation at $10^{-35}$, we yield $y=46.7558\approx4.75\pi^2$. This illustrates that the divergence of the condition causes the solution for $y$ to grow with increasing truncation accuracy. Since $T^*\propto y$, this implies that the effective temperature becomes arbitrarily large (perhaps unphysically so, even for effective standards) as the minimization becomes more precise. 

Using our two truncation examples as our naive boundaries, and using the lattice constant $a=2.707\,l_P$, we obtain the effective temperature at its lower and upper bounds corresponding to the respective values of $y$:
\begin{eqnarray}
&T_\text{lower}^*\approx 2.3211~{m_P}/{k_B}\approx 0.739\pi~T_P,\nonumber\\
&T_\text{upper}^*\approx 11.0252~{m_P}/{k_B}\approx1.117\pi^2~T_P,
\end{eqnarray}
where $T_P=1.416\times10^{32}$ K is the Planck temperature. Despite the present day vacuum being much colder than these scales, the gravitons mediating lattice oscillations are so energetically stochastic that they effectively behave as a very hot quantum gas. For reference, the energy scale associating with the Planck temperature is $k_BT_P\simeq1.22\times10^{22}~\text{MeV}$; this is an energy far beyond any standard model process. Using the lower bound of the effective temperature in Eq. (\ref{sho}), the SHO energy in real space becomes, recovering $c$:
\begin{equation}
U_\text{SHO}(r)\approx-40.6042\, m_Pc^2\mathrm{csch}^2\left(\frac{0.739 \pi^2}{l_P} r\right).
\end{equation}

Unlike the classic Hookean profile, in which $U_\text{SHO}\propto r^2$, this expression scales asymptotically as $\sim1/r^2$ for large $r$. This difference arises from the exponential suppression imposed by the hyperbolic cosecant function, $\text{csch}(x)=1/\sinh{(x)}$, where the argument grows rapidly due to the large prefactor $\sim1/l_P$. In the large-$x$ limit, the hyperbolic sine behaves as $\sinh(x)\sim e^x/2$, or equivalently $\mathrm{csch}(x)\approx 2e^{-x}$ as $x\rightarrow\infty$. Assuming that  $e^{-x}\sim1/x$ for large $x$, which mirrors the behavior of exponential decay under truncation, we obtain an approximate yet compact algebraic expression for the SHO energy: 
\begin{equation}\label{fsho}
U_\text{SHO}(r)\approx-3.0531\, m_Pc^2 \left(\frac{l_P}{r}\right)^2.
\end{equation}

\subsection{Total Energy of the Lattice Point}

Combining the $l=1$ structural energy from Eq. (\ref{fstru}) and the SHO energy from Eq. (\ref{fsho}), the vertex Hamiltonian acquires a displacement-dependent eigenvalue. Each vertex interacts with six neighboring lattice points, and we sum the 12-6 Lennard-Jones terms accordingly. The total energy is the expressed in terms of the Planck mass $m_P$, the Planck length $l_P$, and the vertex displacements $r$ and $r_{ij}$, where $r_{ij}$ is the separation between the vertex and one of its six neighbors:
\begin{widetext}
\begin{equation}\label{entot}
E_\text{tot}(r)\simeq 2m_Pc^2\left(3.084-0.8498\left(\frac{l_P}{r}\right)^2+1.9632\sum_{i\neq j=1}^6\left[\left(\frac{1.1743\,l_P}{r_{ij}}\right)^{12}-\left(\frac{1.1743\,l_P}{r_{ij}}\right)^6\right]\right).
\end{equation}
\end{widetext}

Even in the asymptotic limit $r,~r_{ij}\rightarrow\infty$, the zero-point energy $U_\text{ZP}$ remains as a non-vanishing term residual term. This implies that each lattice vertex possesses a well-defined rest energy in the absence of nearby interactions: a quantum mechanical feature arising purely from the Heisenberg uncertainty principle. The presence of neighboring vertices enhances this energy primarily through the structural contributions governed by the 12-6 potential. 

Given Eq. (\ref{entot}), notice how the $l=1$ angular momentum barrier is taken over by the SHO energy from Eq. (\ref{fsho}), leaving behind a negative-value residue from simple harmonic oscillations. For larger $l$, e.g. $l=2$, the barrier will overpower the SHO contribution, assuming the degeneracy of spin-0 gravitons does not depend on $l$ directly. Since we are treating gravitons as the quanta of vertex vibrations in this study, we can assume that $g_s=1$ in the $l=1$ case leads to the generalization $g_s=l$, as typical simple harmonic oscillators have wavefunction solutions that scale with spherical harmonic functions $Y^m_l(\theta,\phi)$, where $l$ drastically influences the function's profile. 

This proposes that the enhancement in angular structure may lead to an enhanced SHO energy, giving spin-0 gravitons an additional degree in freedom in azimuthal number $l$. This is analogous to spin-1 gluons in quantum chromodynamics, which they exhibit intermingling degrees in freedom in polarization (2) and color-dependent configurations (8) to yield $g_s=16$.

\section{Discussion}

\subsection{A Perturbed Hamiltonian}

Suppose the vertex Hamiltonian, whose energy eigenvalue is given in Eq. (\ref{entot}), is perturbed by the influence of SHO gravitons. Recall from Section \ref{3drep} that each vertex is connected to six neighbors: three internal to the reference tetrahedral cell and three external, shared with adjacent cells. The vibrations of neigboring vertices originate from their own structural interactions and propagate through the lattice. This propagation implies that vibrational residue from connected edges, mediated by SHO gravitons, introduces a perturbative correction to the Hamiltonian. These massless excitations contribute additional energy due to their propagation along the spin-network. The perturbed Hamiltonian acting on the top slice is therefore, with $c=1$:
\begin{equation}\label{pert}
\widehat{H'}_\text{t}|\psi\rangle_\text{t}\equiv E_\text{pert.}|\psi\rangle_\text{t}=\left(E_\text{tot}(r)+i\hbar\nabla \right)|\psi\rangle_\text{t}.
\end{equation}
Here, $E_\text{pert.}$ is the total energy including the perturbation. Since gravitons are massless, the Einstein relation $E=|\vec{p}|$ applies, motivating the use of the momentum operator $\hat{p}=-i\hbar\nabla$ as a first-order Ansatz for the perturbative term. However it is important to note that, in principle, a more complete model would involve summing contributions from neighboring momentum operators weighted by a correlation or propagator term.

To solve for the perturbed energy $E_\text{pert.}$, we employ the variational method, using an energy functional $E[\beta]$ with $\beta$ being a characteristic length scale to be optimized. For the trial wavefunction $|\psi\rangle_\text{t}$, we construct an Ansatz combining the localized $l=1$ harmonic oscillator (for the vertex) with a delocalized plane wave, representing the influence of a graviton gas. The resulting normalized wavefunction is:
\begin{equation}\label{ans}
|\psi\rangle_\text{t}\approx\frac{2}{3\sqrt{\pi}}\exp\left(-\frac{r^2}{2\beta^2}\right)H_2\left(\frac{r}{\beta}\right)\exp\left(i\frac{\pi r}{\beta}\right),
\end{equation}
where $H_n(x)$ is the $n$-th Hermite polynomial. For $l=1$, $n=l+1=2$, corresponding to the second excited oscillator state. Since the wavefunction is readily normalized (i.e., $\langle\psi|\psi\rangle=1$), the energy expectation value is $E[\beta]=\langle\psi|\widehat{H'}_\text{t}|\psi\rangle$, where $\widehat{H'}_\text{t}$ is the perturbed Hamiltonian defined in Eq. (\ref{pert}). The explicit derivation of this functional is provided in Appendix \ref{deriv}. The result, expressed in terms of $\beta$ and $l_P$, reads as follows, recovering $c$:
\begin{widetext}
\begin{equation}\label{funct}
\implies E[\beta]=\frac{32\sqrt{\pi}}{9}\frac{\hbar c}{\beta}+12.3731\,m_Pc^2\left(1+0.5511\left(\frac{l_P}{\beta}\right)^2+11.5736\left(\frac{l_P}{\beta}\right)^{6}+1.6440\left(\frac{l_P}{\beta}\right)^{12} \right).
\end{equation} 
\end{widetext}
The $\beta^{-1}$ term originates from the kinetic contribution of graviton-induced perturbations. The remaining terms represent the structural and zero-point contributions from the unperturbed total energy $E_\text{tot}$. For practical purposes, the terms containing $(l_P/\beta)^6$ and $(l_P/\beta)^{12}$ can be neglected due to their smallness in the near-Planckian regime. To determine the optimal value of $\beta$, we minimize the energy functional $E[\beta]$ from Eq. (\ref{funct}) with respect to $\beta$, yielding
\begin{equation}
\partial_\beta E[\beta]=0\implies\beta=-2.1640\,l_P.
\end{equation}
The value $\beta=-2.1640\,l_P$ corresponds to the global minimum of the energy functional. Substituting this value back into $E[\beta]$, we obtain the perturbed energy:
\begin{equation}\label{enturb}
E[\beta]\approx 10.9170\,m_Pc^2.
\end{equation} 

The perturbed energy obtained in Eq. (\ref{enturb}) is approximately $1.77$ times greater than the vertex's zero-point energy $U_\text{ZP}=6.1674\,m_Pc^2$. Assuming the excess energy arises from vibrational modes (the perturbing gravitons) occupying the vertex slice, spanning one unit of the lattice constant $a$, we identify the difference $E[\beta]-U_\text{ZP}$ as the total energy of $N$ gravitons, each contributing a SHO energy from Eq. (\ref{fsho}) evaluated at $r=a$. This yields 
\begin{equation}
4.7496\,m_Pc^2=N\cdot|U_\text{SHO}(a)|\implies \lfloor N\rfloor=11.
\end{equation} 
Thus, each vertex slice contains approximately nine gravitons in its perturbed state. Including the Bose enhancement factor $(N+1)$ due to coherence, we conclude that the vertex effectively hosts twelve spin-0 gravitons. These quanta may be interpreted as a coherent bath of perturbing modes, emergent from the vibrational coupling across a regular tetrahedral spin-network. 

\subsection{Concluding Remarks}

In this work, we have shown that condensed matter methods can be employed to characterize an elementary spin-network from loop quantum gravity, modeled here as a Planck-scale tetrahedral spacetime lattice. While the Wheeler-DeWitt equation enforces a global Hamiltonian constraint on physical states, the ADM formulism enables a decomposition into spatial hypersurfaces, each with a well-defined, non-zero local Hamiltonian. This foliation allows us to analyze localized quantum dynamics within the spin-network. 

We demonstrated that the Hamiltonian associated with a single vertex slice can be constructed using the Schr\"odinger equation. The vertex Hamiltonian includes contributions from rotational kinetic energy (via the angular momentum barrier), cohesive energy modeled through the Lennard-Jones potential and a zero-point quantum term, and vibrational energy from quantized oscillations. Together, these contributions form the structural (Eq. [\ref{fstru}]) and the simple harmonic oscillator energy (Eq. [\ref{fsho}]), which collectively describe the energy eigenvalue of a lattice point in this quantum spacetime model.
 
Quantizing the SHO energy using a free gas model for spin-0 gravitons (whose Hamiltonian is  $\widehat{H}=\widehat{p}c$), we determine the energy eigenvalue of a perturbed vertex via the variational method. The resulting eigenvalue (Eq. [\ref{enturb}]) indicates that the vertex slice, and by extension the full tetrahedral lattice, is densely populated with free graviton-like excitations, modeled here as spin-0 pseudoparticles. This further supposes that gravitons travel across distances smaller than the Planck length (see e.g. Ref. \cite{MacKay:2024qxj}).

Since this study focused on the Planck-scale tetrahedral lattice as the structural unit of unperturbed (flat) spacetime, a natural extension would be to analyze a unit of ``bent'' spacetime. In such a scenario, the Wheeler-DeWitt equation would generalize into a Schr\"odinger-like equation with $\widehat{\mathcal{H}}|\Psi\rangle\neq0$, and the spin values of the six edges would shift toward integer (bosonic) values, reflecting curvature-induced modifications to area quantization. It is conjectured that, in this regime, the total area of the spin-network approaches the surface area of a quantum black hole, $A_\text{QBH}=16\pi l_P^2$. If the Brownian graviton gas is re-analyzed in this context, it may provide a mechanism for testing proposals that quantum black holes (or geometric centers of larger black holes) function as graviton reservoirs (as proposed as a thought experiment in Ref. \cite{Kimura:2023tde} and further discussed in Ref. \cite{MacKay:2024qxj}). 


\begin{appendix}

\section{Length and Height Calculations}\label{calc}

The tetrahedral cell, depicted in Figure \ref{fig:sub}, is assumed to be composed of equilateral triangular faces. To compute the height $h$ from the apex to the centroid of the opposite face, we solve two right-triangular relations using the Pythagorean theorem. 

First, consider the base face. For an equilateral triangle, the medians from each corner have the same length $b$, where the centroid of the face is marked at the half distance $b/2$. Dividing a equilateral triangle with a median forms two right triangles, where each right triangle has a horizontal leg of length $a/2$ and a hypotenuse of length $a$. Note that $a$ is the lattice constant (the edge length). Solving for the vertical leg $b$, which is the full length of the median, we find:
\begin{equation}
\left(\frac{a}{2}\right)^2+b^2=a^2\implies b=a\frac{\sqrt{3}}{2}.
\end{equation} 
The centroid is therefore marked at $b/2=a\sqrt{3}/4$.

Next, consider a right triangle containing the base face centroid, a corner vertex and the apex. The horizontal leg of the triangle is now $b/2$, and the vertical leg is the tetrahedral height $h$; the hypotenuse remains $a$. Applying the Pythagorean theorem again:
 \begin{equation}
\left(\frac{b}{2}\right)^2+h^2=a^2\implies h=a\frac{\sqrt{13}}{4}.
\end{equation}

\section{Deriving $E[\beta]$}\label{deriv}

 Given the perturbed Hamiltonian from Eq. (\ref{pert}) and the wavefunction Ansatz of Eq. (\ref{ans}), we begin the variational method by computing the expectation value of the energy functional $E[\beta]$. As the Hamiltonian itself is a linear, first-order differential equation, we find the expectation value of the perturbative Hamiltonian $\widehat{H}_{(1)}=i\hbar\nabla$. The expectation value is an infinite integral over $r/\beta$, where the integrand is explicitly defined as follows: 
\begin{widetext}
\begin{equation}
\psi^*(r)\widehat{H}_{(1)}\psi(r)=\frac{16\hbar}{9\pi\beta^6}\exp\left(-\frac{r^2}{\beta^2}\right)\left(\beta^2-2r^2\right)\cdot\mathrm{sgn}(\beta)^4\left[\beta^3\pi-2\beta\pi r^2+i(5\beta^2r-2r^3)\right],
 \end{equation}
 \end{widetext}
 where $\mathrm{sgn}(\beta)^4=1$ ($\mathrm{sgn}(\beta)$ being the signum function of the length scale $\beta$), assuming $\beta$ is a real number. Therefore, the integral evaluation yields the expectation value of the perturbing Hamiltonian: 
 \begin{equation}
\int_{-\infty}^\infty \psi^*(r)\widehat{H}_{(1)}\psi(r)\frac{dr}{\beta}=\frac{32\sqrt{\pi}}{9}\frac{\hbar}{\beta}.
 \end{equation}
 
 The unperturbed Hamiltonian has a well-defined eigenvalue $E_\text{tot}(r)$ from Eq. (\ref{entot}), though it is explicitly a function of position. Thus, the energy expectation value $\langle\psi|E_\text{tot}(r)|\psi\rangle$ is another infinite integral over $r/\beta$. Due to the presence of divergent terms in $E_\text{tot}(r)$ arising from inverse even powers of $r$ (e.g., $r^{-2}$, $r^{-6}$, $r^{-12}$), we regularize the integral using a truncation method: $r^{-2n}\rightarrow(r^2-\xi^2)^{-n}$ for integer $n\geq1$, and take the limit $\xi\rightarrow0$ via Taylor expansion, isolating the finite, $\xi$-independent contributions. This regulator approach mirrors analytic continuation techniques used in zeta-function regularization and in quantum field theory, but is here tailored to retain only the dominant physically meaningful contributions. This gives the regulated expectation value:
\begin{equation}
\lim_{\xi\rightarrow0}\langle\psi|\widehat{H}_{(0)}|\psi\rangle=\lim_{\xi\rightarrow0}\int_{-\infty}^\infty E_\text{tot}(r,\xi)\psi^*(r)\psi(r)\frac{dr}{\beta}\nonumber
\end{equation}
\begin{widetext}
\begin{equation}
= 12.3731\,m_Pc^2\left(1+0.5511\left(\frac{l_P}{\beta}\right)^2+11.5736\left(\frac{l_P}{\beta}\right)^{6}+1.6440\left(\frac{l_P}{\beta}\right)^{12} \right)+\mathcal{O}(\xi^{-2n})~\forall~n\geq1.
\end{equation}
Neglecting divergent regulator-dependent terms, we retain only the finite part. Together with the perturbative contribution derived earlier, we obtain the full energy functional as Eq. (\ref{funct}), recovering $c$:
\begin{equation}
 E[\beta]=\frac{32\sqrt{\pi}}{9}\frac{\hbar c}{\beta}+12.3731\,m_Pc^2\left(1+0.5511\left(\frac{l_P}{\beta}\right)^2+11.5736\left(\frac{l_P}{\beta}\right)^{6}+1.6440\left(\frac{l_P}{\beta}\right)^{12} \right).
\end{equation} 
\end{widetext}

\end{appendix}

\end{document}